\documentclass[a4paper, 10pt,conference]{IEEEtran}

\usepackage{epsfig}
\usepackage{graphicx}
\usepackage{epstopdf}
\usepackage{amsmath}
\usepackage{amsfonts}
\usepackage{amssymb}
\usepackage{setspace}
\begin{document}
\title{Achieving Shannon Capacity Region as Secrecy Rate Region in a Multiple Access Wiretap Channel}
\author{\IEEEauthorblockN{Shahid Mehraj Shah and Vinod Sharma}
\IEEEauthorblockA{Department of Electrical Communication Engineering,\\Indian Institute of Science, Bangalore, India\\
Email: \{shahid, vinod\}@ece.iisc.ernet.in}}
\vspace{1cm}\maketitle

\begin{abstract}
We consider a two user multiple-access channel with an eavesdropper at the receiving end. We use previously transmitted messages as a key in the next slot till we achieve the capacity region of the usual Multiple Access Channel (MAC). 
\end{abstract}


\vspace{0.01cm}
\allowdisplaybreaks
\section{Introduction}
We consider a multiple access wiretap channel (MAC-WT), which is a multiple access channel (MAC) with an eavesdropper tapping the channel. A MAC is one of the fundamental building blocks in multihop networks and also models the uplink of the most common wireless networks, cellular networks (\cite{cover2012elements}).

For a MAC-WT in addition to reliability, the security of a transmitted message is also very important. The security aspect of MAC at the physical layer has been studied for two models of MAC: (i) each transmitting user receives noisy outputs of other users and the users treat each other as eavesdroppers (\cite{liang2008multiple}), (ii) the eavesdropper is at the receiving end (\cite{tekin2008gaussian}). In both the cases, as in a single user case (\cite{wyner1975wire}), due to security constraints, there is a loss of achievable rates i.e., the rate region shrinks due to security constraints on each user \cite{liang2008multiple}, \cite{liang2009information}, \cite{tekin2008gaussian}. In other multiuser scenarios also there is trade-off between secrecy and the achievable rate region \cite{liang2009capacity}, \cite{el2013achievable}, \cite{oohama2007capacity}.

A Fading MAC with an eavesdropper was studied in \cite{tekin2007secrecy}, where the authors assume that complete channel state of the eavesdropper is available at the transmitter. In \cite{shah2012achievable} the authors showed that even if no CSI of the eavesdropper is available at the transmitter (only the distribution of channel states is known) it is still possible to achieve positive secrecy rates.

In a single user wiretap channel there have been various efforts to improve secrecy rate. Most of this work relies on a public channel or feedback. The public channel is used to exchange a function of some correlated random variables (which the transmitter and receiver have access to before the communication starts), and at the end of the protocol the transmitter and receiver agree on a secret key about which the eavesdropper (Eve) has very little knowledge \cite{ahlswede1993common}, \cite{maurer1993secret}. This key can then be used along with a stochastic encoder to enhance the overall secrecy rate \cite{yamamoto1997rate}, \cite{kang2010wiretap}. In the other case a feedback channel from the main receiver to the transmitter is used to exchange the secret key, which can be used to enhance the secrecy rate \cite{ardestanizadeh2009wiretap}, \cite{lai2008wiretap}. 

In some cases the secrecy capacity equal to the main channel capacity is achieved. In \cite{kobayashi2013secure} the authors propose a simplex coding scheme to achieve ordinary channel capacity in a wiretap channel. In this paper there are multiple messages to be transmitted. The authors are able to show that only the currently transmitted message is secure with respect to (w.r.t.) the outputs of the eavesdropper.

In \cite{shah2013previous} the authors proposed a simple coding scheme in which the previous securely transmitted messages are used as secret keys in the future slots along with a stochastic encoder to achieve the secrecy capacity close to the main channel capacity. In this paper also, the authors are able to show that the messages transmitted in a slot are secure w.r.t. all the previous outputs of the eavesdropper. This scheme was strengthened in \cite{shah2014achieving} in which the legitimate users have a secret key buffer to store the secret messages. The authors prove that not only the current message but all the messages transmitted after a particular slot are secure w.r.t. all the information that the eavesdropper has till the present slot.

In this paper we extend the coding schemes of \cite{shah2013previous} and \cite{shah2014achieving} to a MAC-WT and prove that via this scheme we can achieve the whole capacity region of the MAC. Although in the paper due to lack of space we show the secrecy of individual messages as in \cite{shah2013previous}, the coding scheme permits us to show the secrecy achieved in \cite{shah2014achieving} and will be shown in a future work.

The rest of the paper is organised as follows. In Section II we introduce the channel model and formulate the problem. In Section III we present the main result. Finally we conclude the paper in Section IV. Proof of a lemma is provided in the Appendix.

\section{Multiple Access Wiretap Channel}
We consider a discrete time, memoryless two user MAC-WT $(\mathcal{X}_1\times \mathcal{X}_2,p(y,z|x_1,x_2),\mathcal{Y},\mathcal{Z})$, where $\mathcal{X}_1,\mathcal{X}_2,\mathcal{Y},\mathcal{Z}$ are finite sets and $p(y,z|x_1,x_2)$ is the collection of conditional probability mass functions characterizing the channel. We denote by $X_1\in \mathcal{X}_1$ and $X_2\in \mathcal{X}_2$, the inputs from the two users to the channel, and by $Y\in \mathcal{Y}$ the corresponding output of the channel received by Bob, the legitimate receiver and $Z\in \mathcal{Z}$ the output received by Eve. The two users want to send messages $W^{(1)}$ and $W^{(2)}$ to Bob reliably, while keeping Eve ignorant about the messages.

\textbf{Definition}: For a MAC-WT, a $(2^{nR_1},2^{nR_2},n)$ codebook consists of (1) Message sets $\mathcal{W}^{(1)}$ and $\mathcal{W}^{(2)}$ of cardinality $2^{nR_1}$ and $2^{nR_2}$, (2) Messages $W^{(1)}$ and $W^{(2)}$, which are uniformally distributed over the corresponding message sets $\mathcal{W}^{(1)}$ and $\mathcal{W}^{(2)}$ and are independent of each other, (3) two stochastic encoders, 
\begin{equation}
f_i:\mathcal{W}^{(i)} \rightarrow \mathcal{X}_i^n~~i=1,2,
\end{equation}
and 4) Decoder at Bob: 
\begin{equation}
g:\mathcal{Y}^n\rightarrow \mathcal{W}^{(1)} \times \mathcal{W}^{(2)}.
\end{equation}
where  $(Y_1,\ldots,Y_n) \triangleq Y^n \in \mathcal{Y}^n$. The decoded messages are denoted by $(\widehat{W}^{(1)},\widehat{W}^{(2)})$.

The average probability of error at Bob is 
\begin{equation}
P_e^{(n)}\triangleq P\left\lbrace \left(\widehat{W}^{(1)},\widehat{W}^{(2)}\right) \neq \left(W^{(1)},W^{(2)}\right) \right\rbrace.
\end{equation}

\textit{Leakage Rate} is 
\begin{equation}
R_L^{(n)}=\frac{1}{n}I(W^{(1)},W^{(2)};Z^n),
\end{equation}
the rate at which information is leaked to Eve. We consider perfect secrecy capacity rates.


\textit{Definition}: The secrecy-rates $(R_1,R_2)$ are achievable if there exists a sequence of codes $(2^{nR_1},2^{nR_2},n)$ with $P^{(n)}_e\rightarrow 0$ as $n\rightarrow \infty$ and
\begin{align}
\limsup_{n\rightarrow \infty} R_{L}^{(n)} =0.
\end{align}
The secrecy-rate region is the closure of the convex hull of achievable secrecy-rate pair $(R_1,R_2)$.


In \cite{tekin2008gaussian}, \cite{tekin2010correction} a coding scheme to obtain the following rate region was proposed.

\begin{figure}[!htb]
\centering
\includegraphics[scale=0.8]{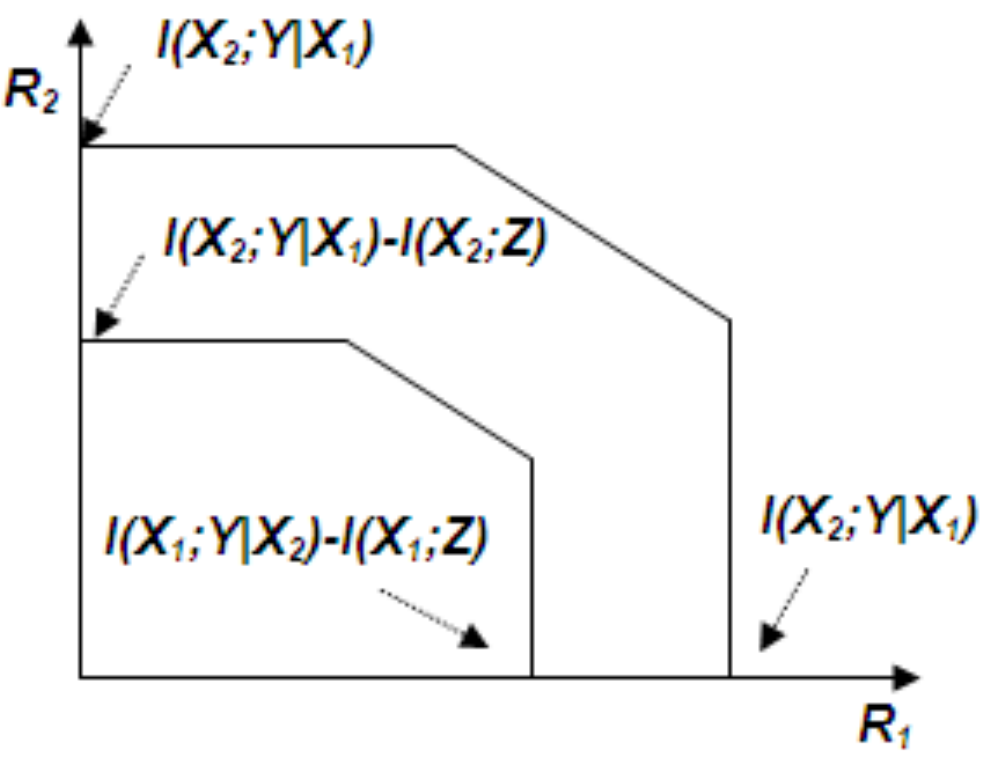}
\caption{Achievable rates for MAC-WT with and without secrecy constraints}
\label{fig:digraph}
\end{figure}

\textit{Theorem 1}: Rates $(R_1,R_2)$ are achievable with $\limsup_{n\rightarrow \infty}R_L^{(n)}=0$, if there exist independent random variables $(X_1,X_2)$ satisfying
\begin{align}
R_1&\leq I(X_1;Y|X_2)-I(X_1;Z),  \nonumber \\
R_2&\leq I(X_2;Y|X_1)-I(X_2;Z),  \nonumber \\
R_1+R_2&\leq I(X_1,X_2;Y)-I(X_1;Z)-I(X_2;Z). ~~ \square  \label{MAC_WT_Secrecy_region}
\end{align}
\quad The secrecy capacity region for a MAC-WT is not known. If the secrecy constraint is not there then the capacity region for MAC obtained from the convex closure of the regions in Theorem 1 without the term $I(X_i;Z),~i=1,2$ on the right side of (\ref{MAC_WT_Secrecy_region}) (\ref{fig:digraph}). In the next section we show that we can attain the capacity region of a MAC even when some secrecy constraints are satisfied.

\section{Enhancing the Secrecy-Rate Region of MAC-WT}
In this section we extend the coding-decoding scheme of \cite{shah2013previous} for a point-to-point channel to enhance the achievable secrecy rates for a MAC-WT. We recall that in \cite{shah2013previous} the system is slotted with a slot consisting of $n$ channel uses. The first message is transmitted by using the wiretap code of \cite{wyner1975wire} in slot 1. In the next slot we use the message transmitted in slot 1 as a key along with wiretap code and transmit two messages in that slot (keeping the number of channel uses same). Hence the secrecy-rate gets doubled. We continue to use the message transmitted in the previous slot as a key and wiretap coding, increasing the transmission rate till we achieve a secrecy rate equal to the main channel capacity. From then onwards we use only the previous message as key and no wiretap coding. This scheme guarantees that the message which is currently being transmitted is secure w.r.t. all the Eavsdropper's outputs, i.e., if message $W_k$ is transmitted in slot $k$ then 
\begin{equation}
\frac{1}{n}I(W_k;Z_1^n,\ldots,Z_k^{n})\rightarrow 0,  \label{leakage_slot_k_1}
\end{equation}
as the codeword length $n \rightarrow \infty$, where $Z^n_i$ is the data received by Eve in slot $i$.

In the following, not only we extend this coding scheme to a MAC-WT but also modify it so that it can be used to improve its secrecy criterion (\ref{leakage_slot_k_1}) and for fading channels as well. These extensions will be presented in future work. Currently the secrecy criterion used in the following is: If user $i$ transmits message $\overline{W}_k^{(i)}$ in slot $k$, we need
\begin{equation}
I(\overline{W}_l^{(1)},\overline{W}_l^{(2)};Z_1^n,\ldots,Z_k^{n}) \leq 2n_1 \epsilon,~\text{for}~l=1,\ldots,k \label{leakage_slot_k_induction}
\end{equation}
for any given $\epsilon>0$. This will be strengthened to strong secrecy, $I(\overline{W}_k^{(1)},\overline{W}_k^{(2)};Z_1^n,\ldots,Z_k^{n}) \rightarrow 0~\text{as}~n\rightarrow \infty$ at the end of the section. We modify message sets and encoders and decoders with respect to Section II as follows.

 The message sets are $\mathcal{W}^{(i)}=\{1,\ldots,2^{nR^s_i}\}$ for users $i=1,2$, where $(R^s_1,R^s_2)$ satisfy (\ref{MAC_WT_Secrecy_region}) for some $(X_1,X_2)$. The encoders have two parts for both users,
\begin{equation}
f^s_1:\mathcal{W}^{(1)}\rightarrow \mathcal{X}_1^{n_1}, ~~ f^d_1:\mathcal{W}^{(1)}\times \mathcal{K}_1\rightarrow \mathcal{X}_1^{n_2}
\end{equation}
\begin{equation}
f^s_2:\mathcal{W}^{(2)}\rightarrow \mathcal{X}_2^{n_1}, ~~ f^d_2:\mathcal{W}^{(2)}\times \mathcal{K}_2\rightarrow \mathcal{X}_2^{n_2},
\end{equation}
where $X_i\in \mathcal{X}_i, i=1,2$, and $\mathcal{K}_i, i=1,2$ are the sets of secret keys generated for the respective user, $f^s_i, i=1,2$ are the wiretap encoders corresponding to each user as in \cite{tekin2008gaussian} and $f_i^d, i=1,2$ are the usual deterministic encoders corresponding to each user . Also, $n_1+n_2=n$.

Observe that, for transmitted messages $W^{(i)},i=1,2$ and the corresponding received data $Z^n$ by Eve,
\begin{align}
I(&W^{(1)},W^{(2)};Z^n)  \nonumber \\
&=I(W^{(1)};Z^n)+I(W^{(2)};Z^n\lvert W^{(1)}) \nonumber \\
&=H(W^{(1)})-H(W^{(1)}\lvert Z^n)+H(W^{(2)})  \nonumber \\
&\quad-H(W^{(2)}\lvert Z^n,W^{(1)}) \nonumber \\
&\overset{(a)}{\leq}H(W^{(1)}\lvert X_2^n)-H(W^{(1)}\lvert Z^n,X_2^n)+H(W^{(2)}\lvert X_1^n) \nonumber \\
&\quad -H(W^{(2)}\lvert Z^n,X_1^n) \nonumber \\
&=I(W^{(1)};Z^n\lvert X_2^n)+I(W^{(2)};Z^n\lvert X_1^n),  \label{leakage_2}
\end{align}
where $(a)$ follows from the facts: conditioning decreases entropy, messages are independent and a codeword is a function of the message to be transmitted. Thus if $\limsup_{n\rightarrow \infty} I(W^{(i)};Z^n\lvert X_j^n)/n= 0, i=1,2$, then $R_L^{(n)}\rightarrow 0$.

The following is our main result.

\textit{Theorem 2}: The secrecy-rate region satisfying (\ref{leakage_slot_k_induction}) is the usual MAC region without Eve, i.e., it is the closure of convex hull of all rate pair $(R_1,R_2)$ satisfying
\begin{align}
R_1&\leq I(X_1;Y|X_2), R_2\leq I(X_2;Y|X_1),  \nonumber \\
R_1+R_2&\leq I(X_1,X_2;Y), \label{MAC_capacity_region}
\end{align}
for some independent random variables $X_1,X_2$.

\textit{Proof}: We fix distributions $p_{X_1},p_{X_2}$. Initially we take $n_1=n_2=n/2$. In slot 1, user $i$ selects message $\mathcal{W}_1^{(i)} \in \mathcal{W}^{(i)}$ to be transmitted confidentially. Both the users use the wiretap coding scheme of \cite{tekin2008gaussian}. Hence the rate pair $(R_1,R_2)$ satisfies (\ref{MAC_WT_Secrecy_region}) and (\ref{leakage_slot_k_1}).
In slot 2, each user uses previously transmitted secure message as key to enhance the rate. The two users select two messages each to be transmitted, $(\overline{W}^{(1)}_{21},\overline{W}^{(1)}_{22})\equiv \overline{W}^{(1)}_2$ and $(\overline{W}^{(2)}_{21},\overline{W}^{(2)}_{22})\equiv \overline{W}^{(2)}_2$. Both users use the wiretap coding scheme (as in \cite{tekin2008gaussian}) for first  part of the message, i.e., $(\overline{W}^{(1)}_{21},\overline{W}^{(2)}_{21})$, and for the second part user $i$ first takes $XOR$ of $\overline{W}^{(i)}_{22}$ with the previous message, i.e., $\overline{W}^{(i)}_{22}\oplus \overline{W}^{(i)}_1$. This $XOR$ed message is transmitted over the MAC-WT using a usual MAC coding scheme (\cite{ahlswede1973multi}, \cite{liao1972multiple}). Hence the secure rate achievable in both parts of slot 2 satisfies (\ref{MAC_WT_Secrecy_region}) for both the users. This is also the overall rate of slot 2.

In slot 3, in the first part the rate satisfies (\ref{MAC_WT_Secrecy_region}) via wiretap coding. But in the second part we $XOR$ with $\overline{W}_2^{(i)}$ and are able to send two messages and hence \textit{double} the rate of (\ref{MAC_WT_Secrecy_region}) (assuming $2(R_1,R_2)$ via (\ref{MAC_WT_Secrecy_region}) is within the range of (\ref{MAC_capacity_region})). We continue like this.

Define
\begin{equation}
\lambda_1 \triangleq \left\lceil\frac{I(X_1;Y\lvert X_2)}{I(X_1;Y\lvert X_2)-I(X_1;Z)}\right\rceil,  \label{lambda_A_1}
\end{equation}
where $\lceil x\rceil$ is the smallest integer $\geq x$. In slot $\lambda_1+1$ the rate of user 1 in second part of the slot satisfies,
\begin{align}
R_1 & \leq \min \left( \lambda_1 \left( I(X_1;Y\lvert X_2)-I(X_1;Z) \right), I(X_1;Y\lvert X_2)\right) \nonumber \\
&=I(X_1;Y\lvert X_2).
\end{align}
Similarly we define $\lambda_2$ as
\begin{equation}
\lambda_2 \triangleq \left \lceil\frac{I(X_2;Y|X_1)}{I(X_2;Y|X_1)-I(X_2;Z)} \right \rceil. \label{lambda_A_1}
\end{equation}
In slot $\lambda_2+1$, the rate $R_2$ will satisfy
\begin{align}
R_2 \leq I(X_2;Y|X_1).
\end{align}
In slot $\lambda=\max\{\lambda_1,\lambda_2\}+1$, the sum-rate will satisfy
\begin{align}
R&_1+R_2 \leq  \\ \nonumber
& \min \left\lbrace \lambda \left[I(X_1,X_2;Y)-\sum_{i=1}^2 I(X_i;Z)\right], I(X_1,X_2;Y)\right\rbrace.
\end{align}
After some slot, say, $\lambda^*>\lambda$, the sum-rate will get saturated by sum-capacity term, i.e., $I(X_1,X_2;Y)$, and hence thereafter the rate pair $(R_1,R_2)$ in the second part of the slot will satisfy (\ref{MAC_capacity_region}) and the overall rate for the slot is the average in the first part and the second part of the slot.

In slot $k$, (where $k>\lambda^*$) to transmit a message pair $(\overline{W}^{(1)}_k, \overline{W}^{(2)}_k)$, where $\overline{W}^{(i)}_k=(\overline{W}^{(i)}_{k,1},\overline{W}^{(i)}_{k,2}), i=1,2$, we use wiretap coding for $(\overline{W}^{(1)}_{k,1},\overline{W}^{(2)}_{k,1})$ and for the second part, we $XOR$ it with the previous message i.e., $\overline{W}^{(i)}_{k,2}\oplus \overline{W}^{(i)}_{k-1}, i=1,2$, and transmit the overall codeword over the MAC-WT.

 To get the overall rate of a slot close to that in (\ref{MAC_capacity_region}), we make $n_2=ln_1$. By taking $l$ large enough, we can come arbitrarily close to the region in (\ref{MAC_capacity_region}). For this coding scheme, $P_e^n\rightarrow 0$.  Combination of the rates in (\ref{MAC_capacity_region}) can be obtained by time sharing. Now we show that, it also satisfies (\ref{leakage_slot_k_induction}).
\\
\textit{Leakage Rate Analysis}: 
Before we proceed, we define the notation to be used here. For user $i$, the codeword sent in slot $k$ will be represented by $\textbf{X}^{(i)}_k$. Correspondingly $\textbf{X}^{(i)}_{k,1}$ and $\textbf{X}^{(i)}_{k,2}$ will represent $n_1$-length and $n_2$-length codewords of user $i$ in slot $k$. When we consider $i$ to be 1 or 2, $\bar{i}$ will be taken as 2 or 1 respectively. In slot $k$, the noisy codeword received by Eve is $Z_k^{n}\equiv (Z_{k,1}^{n_1},Z_{k,2}^{n_2})$, where $Z_{k,1}^{n_1}$ is the sequence corresponding to the wiretap coding part and $Z_{k,2}^{n_2}$ is corresponding to the $XOR$ part (in which the previous message is used as a key). We also note that in slot 1, $n=n_1$ (i.e., it does not have the second part). For random variables $X,Y$, $X\perp Y$ denotes that $X$ is independent of $Y$.

In slot 1, since wiretap coding of \cite{tekin2008gaussian} is employed, the leakage rate satisfies
\begin{equation}
I(\overline{W}^{(1)}_1;Z_1^n\lvert \textbf{X}_1^{(2)})\leq n_1\epsilon,~I(\overline{W}^{(2)}_1;Z_1^n\lvert \textbf{X}_1^{(1)})\leq n_1\epsilon.
\end{equation}

For slot 2 we show, for user 1,
\begin{align}
I(\overline{W}^{(1)}_1;Z_1^n,Z_2^{n}\lvert \textbf{X}^{(2}_2) &\leq n_1\epsilon, \nonumber \\
I(\overline{W}^{(1)}_2;Z_1^n,Z_2^{n}\lvert \textbf{X}^{(2)}_2) &\leq n_1\epsilon. 
\end{align}
Similarly one can show for user 2.
We first note that
\begin{align}
I(&\overline{W}^{(1)}_1;Z_1^n,Z_2^{n}\lvert \textbf{X}^{(2)}_2) \nonumber \\
&=I(\overline{W}^{(1)}_1;Z_1^n)+I(\overline{W}^{(1)}_1;Z_2^{n}\lvert Z_1^n,\textbf{X}^{(2)}_2) \nonumber \\
&\overset{(a)}{\leq} n_1\epsilon+H(\overline{W}^{(1)}_1\lvert Z_1^n,\textbf{X}^{(2)}_2)-H(\overline{W}^{(1)}_1\lvert Z_1^n,\textbf{X}^{(2)}_2,Z_2^{n}) \nonumber \\
&\overset{(b)}{=} n_1\epsilon+H(\overline{W}^{(1)}_1\lvert Z_1^n)-H(\overline{W}_1^{(1)}\lvert Z_1^n)=n_1\epsilon.   \label{slot_2_1}
\end{align}
where $(a)$ follows from wiretap coding and $(b)$ follows by the fact that $\textbf{X}_2^{(2)} \perp (\overline{W}_1^{(1)},Z_1^n)$, and $(\textbf{X}^{(2)}_2,Z_2^{n}) \perp (\overline{W}_1^{(1)},Z_1^n)$.

Next consider 
\begin{align}
I(&\overline{W}_2^{(1)};Z_1^n,Z_2^{n}\lvert \textbf{X}_2^{(2)})  \nonumber \\
&=I(\overline{W}_{2,1}^{(1)},\overline{W}_{2,2}^{(1)};Z_1^n,Z_2^{n}\lvert \textbf{X}_2^{(2)})  \nonumber \\
&=I(\overline{W}_{2,1}^{(1)};Z_1^n,Z_2^{n}\lvert \textbf{X}_2^{(2)})+I(\overline{W}_{2,2}^{(1)};Z_1^n,Z_2^{n}\lvert \textbf{X}_2^{(2)},\overline{W}_{2,1}^{(1)})  \nonumber \\
&\triangleq I_1+I_2.  \label{leakage_slot_2}
\end{align}
We get upper bounds on $I_1$ and $I_2$. The first term,
\begin{align}
I_1&=I(\overline{W}_{2,1}^{(1)};Z_1^n,Z_2^{n}\lvert \textbf{X}_2^{(2)}) \nonumber \\
&=I(\overline{W}_{2,1}^{(1)};Z_1^n,Z_{2,1}^{n_1},Z_{2,2}^{n_2}\lvert \textbf{X}_2^{(2)}) \nonumber \\
&=I(\overline{W}_{2,1}^{(1)};Z_{1}^{n}\lvert \textbf{X}_2^{(2)})+I(\overline{W}_{2,1}^{(1)};Z_{2,1}^{n_1} \lvert \textbf{X}_2^{(2)},Z_1^n) \nonumber \\ 
&\quad+I(\overline{W}_{2,1}^{(1)};Z_{2,2}^{n_2} \lvert \textbf{X}_2^{(2)},Z_1^n,Z_{2,1}^{n_1})  \nonumber \\
&\overset{(a)}{=}0+I(\overline{W}_{2,1}^{(1)};Z_{2,1}^{n_1} \lvert \textbf{X}_{2,1}^{(2)},\textbf{X}_{2,2}^{(2)}, Z_1^n) \nonumber \\
&\quad+I(\overline{W}_{2,1}^{(1)};Z_{2,2}^{n_2} \lvert \textbf{X}_2^{(2)},Z_1^n,Z_{2,1}^{n_1}) \nonumber \\
&\triangleq I_{11}+I_{12},  \label{leakage_slot2_P1}
\end{align}
where $(a)$ follows because $Z_1^n \perp (\overline{W}_{21}^{(1)},\textbf{X}_2^{(2)})$. Furthermore,
\begin{align}
I_{11}&=I(\overline{W}_{2,1}^{(1)};Z_{2,1}^{n_1} \lvert \textbf{X}_{2,1}^{(2)},\textbf{X}_{2,2}^{(2)}, Z_1^n) \nonumber \\
&=H(\overline{W}_{2,1}^{(1)};\lvert \textbf{X}_{2,1}^{(2)},\textbf{X}_{2,2}^{(2)}, Z_1^n) \nonumber \\
&\quad-H(\overline{W}_{2,1}^{(1)};\lvert Z_{2,1}^{n_1},\textbf{X}_{2,1}^{(2)},\textbf{X}_{2,2}^{(2)}, Z_1^n) \nonumber \\
&\overset{(a)}{=}H(\overline{W}_{2,1}^{(1)};\lvert \textbf{X}_{2,1}^{(2)})-H(\overline{W}_{2,1}^{(1)};\lvert Z_{2,1}^{n_1},\textbf{X}_{2,1}^{(2)}) \nonumber \\
&=I(\overline{W}_{2,1}^{(1)};Z_{2,1}^{n_1},\lvert \textbf{X}_{2,1}^{(2)}) \overset{(b)}{\leq}n_1\epsilon,  \label{leakage_slot_P2}
\end{align}
where $(a)$ follows since $(\textbf{X}_{2,2}^{(2)}, Z_1^n) \perp (\overline{W}_{2,1}^{(1)}, Z_{2,1}^{n_1},\textbf{X}_{2,1}^{(2)})$ and $(b)$ follows since the first part of the message is encoded via the usual coding scheme for MAC-WT. 

Also,
\begin{align}
I_{12}&=I(\overline{W}_{2,1}^{(1)};Z_{2,2}^{n_2} \lvert \textbf{X}_2^{(2)},Z_1^n,Z_{2,1}^{n_1}) \nonumber \\
&=H(\overline{W}_{2,1}^{(1)};\lvert \textbf{X}_{2,1}^{(2)},\textbf{X}_{2,2}^{(2)},Z_1^n,Z_{2,1}^{n_1}) \nonumber \\ 
&\quad -H(\overline{W}_{2,1}^{(1)}\lvert \textbf{X}_{2,1}^{(2)},\textbf{X}_{2,2}^{(2)},Z_1^n,Z_{2,1}^{n_1},Z_{2,2}^{n_2} ) \nonumber \\
&\overset{(a)}{=}H(\overline{W}_{2,1}^{(1)};\lvert \textbf{X}_{2,1}^{(2)}, Z_{2,1}^{n_1})-H(\overline{W}_{2,1}^{(1)};\lvert \textbf{X}_{2,1}^{(2)},Z_{2,1}^{n_1}) =0,  \nonumber
\end{align}
where $(a)$ follows since $(\textbf{X}_{2,2}^{(2)},Z_1^n,Z_{2,2}^{n_2}) \perp (\overline{W}_{2,1}^{(1)},\textbf{X}_{2,1}^{(2)},Z_{2,1}^{n_1})$.

From (\ref{leakage_slot_2}), (\ref{leakage_slot2_P1}) and (\ref{leakage_slot_P2}) we have $I_1=I_{11}+I_{12}\leq n_1\epsilon$. 

Next consider,
\begin{align}
I_2&=I(\overline{W}_{2,2}^{(1)};Z_1^n,Z_2^{n}\lvert \textbf{X}_2^{(2)},\overline{W}_{2,1}^{(1)})  \nonumber \\
&=I(\overline{W}_{2,2}^{(1)};Z_2^{n}\lvert \textbf{X}_2^{(2)},\overline{W}_{2,1}^{(1)})  \nonumber \\
&\quad+I(\overline{W}_{2,2}^{(1)};Z_1^n\lvert \textbf{X}_2^{(2)},\overline{W}_{2,1}^{(1)},Z_2^{n}).  \label{leakage_slot2_I_2}
\end{align}
We have,
\begin{align}
I&(\overline{W}_{2,2}^{(1)};Z_2^{n}\lvert \textbf{X}_2^{(2)},\overline{W}_{2,1}^{(1)}) \nonumber \\
&=I(\overline{W}_{2,2}^{(1)};Z_{2,1}^{n1}\lvert \textbf{X}_2^{(2)},\overline{W}_{2,1}^{(1)}) \nonumber \\
&\quad +I(\overline{W}_{2,2}^{(1)};Z_{2,2}^{n_2}\lvert \textbf{X}_2^{(2)},\overline{W}_{2,1}^{(1)},Z_{2,1}^{n_1})  \nonumber \\
&\overset{(a_1)}{=}0+I(\overline{W}_{2,2}^{(1)};Z_{2,2}^{n_2}\lvert \textbf{X}_2^{(2)},\overline{W}_{2,1}^{(1)},Z_{2,1}^{n_1}) \nonumber \\
&\overset{(a_2)}{=}I(\overline{W}_{2,2}^{(1)};Z_{2,2}^{n_2}\lvert \textbf{X}_{2,2}^{(2)}) \overset{(a_3)}{=}0,  \nonumber
\end{align}
 and $(a_1)$ follows since $\overline{W}_{2,2}^{(1)} \perp (Z_{2,1}^{n_1},\textbf{X}_2^{(2)},\overline{W}_{2,1}^{(1)})$; $(a_2)$ holds because $(\textbf{X}_{2,1}^{(2)},\overline{W}_{2,1}^{(1)})\perp (\overline{W}_{2,2}^{(1)},Z_{2,2}^{n_1},\textbf{X}_{2,2}^{(2)})$; and $(a_3)$ is true since $\overline{W}_{2,2}^{(1)} \perp (\textbf{X}_{2,2}^{(2)},Z_{2,2}^{n_2})$.

 Also,
 \begin{align}
I&(\overline{W}_{2,2}^{(1)};Z_1^n\lvert \textbf{X}_2^{(2)},\overline{W}_{2,1}^{(1)},Z_2^{n})  \nonumber \\
&=I(\overline{W}_{2,2}^{(1)};Z_1^n\lvert \textbf{X}_{2,1}^{(2)},\textbf{X}_{2,2}^{(2)},\overline{W}_{2,1}^{(1)},Z_{2,1}^{n_1},Z_{2,2}^{n_2}) \nonumber \\
&\overset{(b_1)}{=}I(\overline{W}_{2,2}^{(1)};Z_1^n\lvert \textbf{X}_{2,2}^{(2)},Z_{2,2}^{n_2}) \overset{(b_2)}{=}0,  \nonumber
 \end{align}
 where $(b_1)$ follows since $(\overline{W}_{2,1}^{(1)},Z_{2,1}^{n_1},\textbf{X}_{2,1}^{(2)}) \perp (Z_{2,2}^{n_2},\textbf{X}_{2,2}^{(2)},\overline{W}_{2,2}^{(1)},Z_1^n)$), and $(b_2)$ follows because $Z_1^n\perp (\overline{W}_{2,2}^{(1)},\textbf{X}_{2,2}^{(2)},Z_{2,2}^{n_2})$.
 Hence from (\ref{leakage_slot2_I_2}) we have $I_2=0$.
 
 From (\ref{leakage_slot_2}) we have
\begin{equation}
I(\overline{W}_2^{(1)};Z_1^n,Z_2^{n}\lvert \textbf{X}_2^{(2)}) \leq n_1\epsilon.
\end{equation}
Similarly one can show that 
\begin{equation}
I(\overline{W}_2^{(2)};Z_1^n,Z_2^{n}\lvert \textbf{X}_2^{(1)}) \leq n_1\epsilon.
\end{equation}
Therefore , from (\ref{leakage_2}),
\begin{align}
&I(\overline{W}^{(1)}_2,\overline{W}^{(2)}_2;Z_1^n,Z_2^{n})\nonumber \\
&\leq  I(\overline{W}^{(1)}_2;Z_1^n,Z_{2}^n\lvert \textbf{X}^{(2)}_2)+I(\overline{W}^{(2)}_2;Z_1^n,Z_{2}^{n}\lvert \textbf{X}^{(1)}_2). \nonumber
\end{align}

To prove that (\ref{leakage_slot_k_induction}) holds for any slot, we use mathematical induction in the following lemma. 

\textit{Lemma 1}: Let (\ref{leakage_slot_k_induction}) hold for $k$, then it also holds for $k+1$.
\\
\textit{Proof}: See Appendix.  $\square$
\subsection{A note about strong secrecy}
The notion of secrecy used above is \textit{weak secrecy}, i.e., if message $W$ is transmitted and Eve receives $Z^n$, then $I(W;Z^n)\leq n\epsilon $. \textit{Strong Secrecy} requires that $I(W;Z^n)\leq \epsilon$.
In single user case, if strong secrecy notion is used instead of weak secrecy, the secrecy capacity does not change (\cite{maurer2000information}). The same result has been proved for Multiple Access Channel with wiretapper in \cite{yassaee2010multiple} using the channel resolvability technique. In our coding scheme of Theorem 2 if we use resolvability based coding in slot 1, and in subsequent slots use both resolvability based coding (in the first part of the slot) and the previous message (which is now strongly secure w.r.t. Eve) as key in the second part of the slot, we can achieve the same secrecy-rate region (capacity region of usual MAC without Eve), satisfying the leakage rate
\begin{align}
\limsup_{n\rightarrow \infty}I(\overline{W}_k^{(1)},\overline{W}_k^{(2)};Z_1^n,Z_2^{n},\ldots,Z_k^{n}) = 0,
\end{align}
as $n\rightarrow \infty$, because in the RHS of (\ref{leakage_slot_k_induction}), we can get $\epsilon$ instead of $2n_1\epsilon$.

\section{Conclusions}
In this paper we revisit the secrecy-rate region for  a multiple access wiretap channel. We show that by using the previous message as a key in the next slot we can achieve secrecy-rate region equal to the capacity region of a MAC.

Our coding scheme can be used to enhance the secrecy criteria of (\ref{leakage_slot_k_induction}) to 
\begin{align}
\frac{1}{n}I(\overline{W}^{(1)}_k,\overline{W}^{(2)}_k,\overline{W}^{(1)}_{k-1},\overline{W}^{(2)}_{k-1}, \ldots, \overline{W}^{(1)}_{k-N},\overline{W}^{(2)}_{k-N};\textbf{Z}) \rightarrow 0, \nonumber
\end{align}
as $n\rightarrow\infty$, where $\textbf{Z}=(Z_1^n,\ldots,Z_k^n)$ and $N$ can be taken as large as we wish. This will be shown in a future work.
\appendix
\subsection*{Proof of Lemma 1}
In slot $k+1$, for $m=1,\ldots,k$,
\begin{align}
	I(&\overline{W}_{m}^{(1)};Z_{1}^n,Z_2^{n},\ldots,Z_{k+1}^{n}\lvert \textbf{X}_{k+1}^{(2)}) \nonumber \\
	&=I(\overline{W}_{m}^{(1)};Z_{1}^n,Z_2^{n},\ldots,Z_{k}^{n}\lvert \textbf{X}_{k+1}^{(2)})  \nonumber \\
	&+I(\overline{W}_{m}^{(1)};Z_{k+1}^{n}\lvert \textbf{X}_{k+1}^{(2)},Z_{1}^n,Z_2^{n},\ldots,Z_k^{n})  \nonumber \\
	&\triangleq I_1+I_2.    \label{induction_part1}
	\end{align}
	Consider
	\begin{align}
	I_1&=I(\overline{W}_{m}^{(1)};Z_{1}^n,Z_2^{n},\ldots,Z_{k}^{n}\lvert \textbf{X}_{k+1}^{(2)})  \nonumber \\ 
	&\overset{(a)}{=}I(\overline{W}_{m}^{(1)};Z_{1}^n,Z_2^{n},\ldots,Z_{k}^{n}) \nonumber \\
	&=H(\overline{W}_{m}^{(1)})-H(\overline{W}_{m}^{(1)}\lvert Z_{1}^n,Z_2^{n},\ldots,Z_{k}^{n}) \nonumber \\
	&\overset{(b)}{\leq} H(\overline{W}_{m}^{(1)}\lvert \textbf{X}_k^{(2)})-H(\overline{W}_{m}^{(1)}\lvert Z_{1}^n,Z_2^{n},\ldots,Z_{k}^{n},\textbf{X}_k^{(2)}) \nonumber \\
	&= I(\overline{W}_{m}^{(1)};Z_{1}^n,Z_2^{n},\ldots,Z_{k}^{n}\lvert \textbf{X}_k^{(2)}) \overset{(c)}{\leq} n_1\epsilon, \nonumber
		\end{align}
	where $(a)$ follows from $\textbf{X}_{k+1}^{(2)} \perp (\overline{W}_m^{(1)},Z_1^n,Z_{2}^{n},\ldots,Z_k^{n})$, $(b)$ follows from $\overline{W}_m^{(1)} \perp \textbf{X}_k^{(2)}$ and the fact that conditioning decreases entropy; and $(c)$ follows from the induction hypothesis.

    Next consider
	\begin{align}
	I_2&=I(\overline{W}_{m}^{(1)};Z_{k+1}^{n}\lvert \textbf{X}_{k+1}^{(2)},Z_{1}^n,Z_2^{n},\ldots,Z_k^{n})  \nonumber \\
	&=I(\overline{W}_{m}^{(1)};Z_{k+1,1}^{n_1},Z_{k+1,2}^{n_2}\lvert \textbf{X}_{k+1}^{(2)},Z_{1}^n,Z_2^{n},\ldots,Z_k^{n})  \nonumber \\
	&=I(\overline{W}_{m}^{(1)};Z_{k+1,1}^{n_1}\lvert \textbf{X}_{k+1}^{(2)},Z_{1}^n,Z_2^{n},\ldots,Z_k^{n})  \nonumber \\
	&\quad+I(\overline{W}_{m}^{(1)};Z_{k+1,2}^{n_2}\lvert \textbf{X}_{k+1}^{(2)},Z_{1}^n,Z_2^{n},\ldots,Z_k^{n},Z_{k+1,1}^{n_1})  \nonumber \\
	&=I(\overline{W}_{m,1}^{(1)};Z_{k+1,1}^{n_1}\lvert \textbf{X}_{k+1}^{(2)},Z_{1}^n,Z_2^{n},\ldots,Z_k^{n})  \nonumber \\
	&\quad+I(\overline{W}_{m,2}^{(1)};Z_{k+1,1}^{n_1}\lvert \textbf{X}_{k+1}^{(2)},Z_{1}^n,Z_2^{n},\ldots,Z_k^{n},\overline{W}_{m,1}^{(1)})  \nonumber \\
	&\quad+I(\overline{W}_{m,1}^{(1)};Z_{k+1,2}^{n_2}\lvert \textbf{X}_{k+1}^{(2)},Z_{1}^n,Z_2^{n},\ldots,Z_k^{n},Z_{k+1,1}^{n_1})  \nonumber \\
	&\quad+I(\overline{W}_{m,2}^{(1)};Z_{k+1,2}^{n_2}\lvert \textbf{X}_{k+1}^{(2)},Z_{1}^n,\ldots,Z_k^{n},Z_{k+1,1}^{n_1},\overline{W}_{m,1}^{(1)})  \nonumber \\
	&=I_{21}+I_{22}+I_{23}+I_{24}.    \label{induction_part2}
	\end{align}
	Now we consider each of these terms. Observe that,
	\begin{align}
	I_{21}&=I(\overline{W}_{m,1}^{(1)};Z_{k+1,1}^{n_1}\lvert \textbf{X}_{k+1}^{(2)},Z_{1}^n,Z_2^{n},\ldots,Z_k^{n})  \nonumber \\
	&=H(\overline{W}_{m,1}^{(1)}\lvert \textbf{X}_{k+1}^{(2)},Z_{1}^n,Z_2^{n},\ldots,Z_k^{n}) \nonumber \\
	&\quad-H(\overline{W}_{m,1}^{(1)}\lvert \textbf{X}_{k+1}^{(2)},Z_{1}^n,Z_2^{n},\ldots,Z_k^{n},Z_{k+1,1}^{n_1}) \nonumber \\
	&\overset{(a)}{=}H(\overline{W}_{m,1}^{(1)}\lvert Z_{1}^n,Z_2^{n},\ldots,Z_k^{n}) \nonumber \\
	&\quad-H(\overline{W}_{m,1}^{(1)}\lvert Z_{1}^n,Z_2^{n},\ldots,Z_k^{n},Z_{k+1,1}^{n_1},\textbf{X}_{k+1}^{(2)}) \nonumber \\
	&\overset{(b)}{=}H(\overline{W}_{m,1}^{(1)}\lvert Z_{1}^n,Z_2^{n},\ldots,Z_k^{n}) \nonumber \\
	&\quad-H(\overline{W}_{m,1}^{(1)}\lvert Z_{1}^n,Z_2^{n},\ldots,Z_k^{n}) =0,
	\end{align}
	where $(a)$ follows from $\textbf{X}_{k+1}^{(2)} \perp (\overline{W}_{m,1}^{(1)},Z_1^n,\ldots,Z_k^{n})$, $(b)$ follows since $(Z_{k+1,1},\textbf{X}_{k+1}^{(2)}) \perp (\overline{W}_{m,1}^{(1)},Z_1^n,\ldots,Z_k^{n})$.
	
	Next we observe that
	\begin{align}
	I_{22}&=I(\overline{W}_{m,2}^{(1)};Z_{k+1,1}^{n_1}\lvert \textbf{X}_{k+1}^{(2)},Z_{1}^n,Z_2^{n},\ldots,Z_k^{n},\overline{W}_{m,1}^{(1)})  \nonumber \\
	&=H(\overline{W}_{m,2}^{(1)}\lvert \textbf{X}_{k+1}^{(2)},Z_{1}^n,Z_2^{n},\ldots,Z_k^{n},\overline{W}_{m,1}^{(1)})  \nonumber \\
	&\quad-H(\overline{W}_{m,2}^{(1)}\lvert \textbf{X}_{k+1}^{(2)},Z_{1}^n,Z_2^{n},\ldots,Z_k^{n},\overline{W}_{m,1}^{(1)},Z_{k+1,1}^{n_1})  \nonumber \\
	&\overset{(a)}{=}H(\overline{W}_{m,2}^{(1)})-H(\overline{W}_{m,2}^{(1)})=0,
	\end{align}
	where $(a)$ follows since $\overline{W}_{m,2}^{(1)} \perp (\textbf{X}_{k+1}^{(2)},Z_1^n,\ldots,Z_k^{n},Z_{k+1,1}^n,\overline{W}_{m,1}^{(1)})$.
	
	For the third term in (\ref{induction_part2}), we have
	\begin{align}
	I_{23}&=I(\overline{W}_{m,1}^{(1)};Z_{k+1,2}^{n_2}\lvert \textbf{X}_{k+1}^{(2)},Z_{1}^n,Z_2^{n},\ldots,Z_k^{n},Z_{k+1,1}^{n_1})  \nonumber \\
	&=H(\overline{W}_{m,1}^{(1)}\lvert \textbf{X}_{k+1}^{(2)},Z_{1}^n,Z_2^{n},\ldots,Z_k^{n},Z_{k+1,1}^{n_1})  \nonumber \\
	&\quad-H(\overline{W}_{m,1}^{(1)}\lvert \textbf{X}_{k+1}^{(2)},Z_{1}^n,Z_2^{n},\ldots,Z_k^{n},Z_{k+1,1}^{n_1},Z_{k+1,2}^{n_2})  \nonumber \\
	&\overset{(a)}{=}H(\overline{W}_{m,1}^{(1)}\lvert Z_{1}^n,\ldots,Z_k^{n})-H(\overline{W}_{m,1}^{(1)}\lvert Z_{1}^n,\ldots,Z_k^{n})  =0, \nonumber
	\end{align}
    where $(a)$ follows from the fact that $({X}_{k+1}^{(2)},Z_{k+1,1}^n) \perp (\overline{W}_{m,1}^{(1)},Z_{1}^n,Z_2^{n},\ldots,Z_k^{n})$ and $({X}_{k+1}^{(2)},Z_{k+1,1}^n,Z_{k+1,2}^2) \perp (\overline{W}_{m,1}^{(1)},Z_{1}^n,Z_2^{n},\ldots,Z_k^{n})$. 
    
    Finally,
	\begin{align}
	I_{24}&=I(\overline{W}_{m,2}^{(1)};Z_{k+1,2}^{n_2}\lvert \textbf{X}_{k+1}^{(2)},Z_{1}^n\ldots,Z_k^{n},Z_{k+1,1}^{n_1},\overline{W}_{m,1}^{(1)}) =0, \nonumber
	\end{align}
	because $\overline{W}_{m,2}^{(1)} \perp (\textbf{X}_{k+1}^{(2)},Z_{k+1,1}^{n_1},\overline{W}_{m,1}^{(1)},Z_{k+1,2}^n,Z_{1}^n,\ldots,Z_k^{n})$.
	
	From above results and (\ref{induction_part2}) we have $I_2=I_{21}+I_{22}+I_{23}+I_{24}=0$, so from (\ref{induction_part1}) we have	$I=I_1+I_2 \leq n_1\epsilon$, and hence
	\begin{equation}
	I(\overline{W}_{m}^{(1)};Z_{1}^n,Z_2^{n},\ldots,Z_{k+1}^{n}\lvert \textbf{X}_{k+1}^{(2)})  \leq n_1\epsilon,~m=1,\ldots,k. \nonumber
	\end{equation}
	
	Next we show that
	\begin{equation}
	I(\overline{W}_{k+1}^{(1)};Z_{1}^n,Z_2^{n},\ldots,Z_{k+1}^{n}\lvert \textbf{X}_{k+1}^{(2)}) \leq n_1\epsilon. 
	\end{equation}
	We have
	\begin{align}
	I(&\overline{W}_{k+1}^{(1)};Z_{1}^n,Z_2^{n},\ldots,Z_{k+1}^{n}\lvert \textbf{X}_{k+1}^{(2)}) \nonumber \\
	&=I(\overline{W}_{k+1,1}^{(1)},\overline{W}_{k+1,2}^{(1)};Z_{1}^n,Z_2^{n},\ldots,Z_{k+1}^{n}\lvert \textbf{X}_{k+1}^{(2)}) \nonumber \\
	&=I(\overline{W}_{k+1,1}^{(1)};Z_{1}^n,Z_2^{n},\ldots,Z_{k+1}^{n}\lvert \textbf{X}_{k+1}^{(2)}) \nonumber \\
	&\quad+I(\overline{W}_{k+1,2}^{(1)};Z_{1}^n,Z_2^{n},\ldots,Z_{k+1}^{n}\lvert \textbf{X}_{k+1}^{(2)},\overline{W}_{k+1,1}^{(1)}) \nonumber \\
	&=I(\overline{W}_{k+1,1}^{(1)};Z_{1}^n,Z_2^{n},\ldots,Z_{k}^{n}\lvert \textbf{X}_{k+1}^{(2)}) \nonumber \\
	&\quad+I(\overline{W}_{k+1,1}^{(1)};Z_{k+1}^{n}\lvert Z_{1}^n,Z_2^{n},\ldots,Z_{k}^{n}, \textbf{X}_{k+1}^{(2)}) \nonumber \\
	&\quad+I(\overline{W}_{k+1,2}^{(1)};Z_{1}^n,Z_2^{n},\ldots,Z_{k+1}^{n}\lvert \textbf{X}_{k+1}^{(2)},\overline{W}_{k+1,1}^{(1)}) \nonumber \\
	&=I_1+I_2+I_3,   \label{induction_part3}
	\end{align}
	where $I_1=0$ since $(\overline{W}_{k+1,1}^{(1)}, \textbf{X}_{k+1}^{(2)}) \perp  (Z_{1}^n,Z_2^{n},\ldots,Z_{k}^{n} )$, and
	\begin{align}
	I_2&=I(\overline{W}_{k+1,1}^{(1)};Z_{k+1}^{n}\lvert Z_{1}^n,Z_2^{n},\ldots,Z_{k}^{n}, \textbf{X}_{k+1}^{(2)}) \nonumber \\
	&=I(\overline{W}_{k+1,1}^{(1)};Z_{k+1,1}^{n_1},Z_{k+1,2}^n\lvert Z_{1}^n,Z_2^{n},\ldots,Z_{k}^{n}, \textbf{X}_{k+1}^{(2)}) \nonumber \\
	&=I(\overline{W}_{k+1,1}^{(1)};Z_{k+1,2}^{n_2}\lvert Z_{1}^n,Z_2^{n},\ldots,Z_{k}^{n}, \textbf{X}_{k+1}^{(2)}) \nonumber \\
	&\quad+I(\overline{W}_{k+1,1}^{(1)};Z_{k+1,1}^{n_1}\lvert Z_{1}^n\ldots,Z_{k}^{n}, \textbf{X}_{k+1}^{(2)},Z_{k+1,2}^n) \nonumber \\
	&\overset{(a)}{=}0+I(\overline{W}_{k+1,1}^{(1)};Z_{k+1,1}^{n_1}\lvert Z_{1}^n,\ldots,Z_{k}^{n}, \textbf{X}_{k+1}^{(2)},Z_{k+1,2}^n) \nonumber \\
	&=I(\overline{W}_{k+1,1}^{(1)};Z_{k+1,1}^{n_1}\lvert Z_{1}^n,\ldots,Z_{k}^{n}, \textbf{X}_{k+1,1}^{(2)},\textbf{X}_{k+1,2}^{(2)},Z_{k+1,2}^n) \nonumber \\
	&=H(\overline{W}_{k+1,1}^{(1)}\lvert Z_{1}^n,\ldots,Z_{k}^{n}, \textbf{X}_{k+1,1}^{(2)},\textbf{X}_{k+1,2}^{(2)},Z_{k+1,2}^n) \nonumber \\
	&-H(\overline{W}_{k+1,1}^{(1)}\lvert Z_{1}^n,\ldots,Z_{k}^{n}, \textbf{X}_{k+1,1}^{(2)},Z_{k+1,1}^n,\textbf{X}_{k+1,2}^{(2)},Z_{k+1,2}^n) \nonumber \\
	&\overset{(b)}{=}H(\overline{W}_{k+1,1}^{(1)}\lvert \textbf{X}_{k+1,1}^{(2)})-H(\overline{W}_{k+1,1}^{(1)}\lvert  \textbf{X}_{k+1,1}^{(2)},Z_{k+1,1}^n)  \nonumber \\
	&=I(\overline{W}_{k+1,1}^{(1)};Z_{k+1,1}^n\lvert  \textbf{X}_{k+1,1}^{(2)})  \overset{(c)}{\leq}n_1\epsilon,
	\end{align}
	where $(a)$ follows since $(Z_1^n,Z_2^{n},\ldots,Z_k^{n})\perp (\overline{W}_{k+1,1}^{(1)},Z_{k+1,1}^{n_1},\textbf{X}_{k+1}^{(2)})$ and $\overline{W}_{k+1,1}^{(1)}\perp (\textbf{X}_{k+1}^{(2)}), Z_{k+1,1}^{n_1})$; $(b)$ follows since $(Z_1^n,Z_2^{n},\ldots,Z_k^{n},Z_{k+1,2}^n,\textbf{X}_{k+1,2}^{(2)}) \perp (\overline{W}_{k+1,1}^{(1)},\textbf{X}_{k+1,1}^{(2)})$ and  $(Z_1^n,Z_2^{n},\ldots,Z_k^{n},Z_{k+1,2}^n,\textbf{X}_{k+1,2}^{(2)}) \perp (\overline{W}_{k+1,1}^{(1)},Z_{k+1,1}^{n_1},\textbf{X}_{k+1,1}^{(2)})$ and $(c)$ follows from the coding scheme for MAC-WT (\cite{tekin2008gaussian}). 
	
	Now we evaluate $I_3$. We have
	\begin{align}
	I_3&=I(\overline{W}_{k+1,2}^{(1)};Z_{1}^n,Z_2^{n},\ldots,Z_{k+1}^{n}\lvert \textbf{X}_{k+1}^{(2)},\overline{W}_{k+1,1}^{(1)}) \nonumber \\
	&=I(\overline{W}_{k+1,2}^{(1)};Z_{1}^n,Z_2^{n},\ldots,Z_{k}^{n}\lvert \textbf{X}_{k+1}^{(2)},\overline{W}_{k+1,1}^{(1)}) \nonumber \\
	&\quad+I(\overline{W}_{k+1,2}^{(1)};Z_{k+1}^{n}\lvert \textbf{X}_{k+1}^{(2)},\overline{W}_{k+1,1}^{(1)},Z_{1}^n,Z_2^{n},\ldots,Z_k^{n}) \nonumber \\
	&\overset{(a)}{=}0+I(\overline{W}_{k+1,2}^{(1)};Z_{k+1}^{n}\lvert \textbf{X}_{k+1}^{(2)},\overline{W}_{k+1,1}^{(1)},Z_{1}^n,Z_2^{n},\ldots,Z_k^{n}) \nonumber \\
	&=H(\overline{W}_{k+1,2}^{(1)}\lvert \textbf{X}_{k+1}^{(2)},\overline{W}_{k+1,1}^{(1)},Z_{1}^n,Z_2^{n},\ldots,Z_k^{n}) \nonumber \\
	&\quad-H(\overline{W}_{k+1,2}^{(1)}\lvert \textbf{X}_{k+1}^{(2)},\overline{W}_{k+1,1}^{(1)},Z_{1}^n,Z_2^{n},\ldots,Z_k^{n},Z_{k+1}^{n}) \nonumber \\
	&\overset{(b)}{=}H(\overline{W}_{k+1,2}^{(1)})-H(\overline{W}_{k+1,2}^{(1)})=0,
	\end{align}
	where $(a)$ follows since $\overline{W}_{k+1,2}^{(1)} \perp (Z_{1}^n,Z_2^{n},\ldots,Z_k^{n},\textbf{X}_{k+1}^{(2)},\overline{W}_{k+1,1}^{(1)})$; $(b)$ follows since $\overline{W}_{k+1,2}^{(1)} \perp (Z_{1}^n,Z_2^{n},\ldots,Z_k^{n},Z_{k+1}^{n},\textbf{X}_{k+1}^{(2)},\overline{W}_{k+1,1}^{(1)})$.
	Hence from (\ref{induction_part3}), we have
	\begin{equation}
	I(\overline{W}_{k+1}^{(1)};Z_{1}^n,Z_2^{n},\ldots,Z_{k+1}^{n}\lvert \textbf{X}_{k+1}^{(2)}) \leq n_1\epsilon. ~~~\square
	\end{equation}

\bibliographystyle{IEEEtran}
\bibliography{IEEE_WCNC_2015_MAC}

\end{document}